# Comparing three groups

Jelle J. Goeman[1] and Aldo Solari[2]

1: Department of Biomedical Data Sciences, Leiden University Medical Center, Postbus 9600, 2300 RC Leiden, The Netherlands, j.j.goeman@lumc.nl.

2: Department of Economics, Management and Statistics, University of Milano-Bicocca, Piazza dell'Ateneo Nuovo 1, 20126 Milano, Italy

**Abstract**

*We revisit simple and powerful methods for multiple pairwise comparisons that can be used in designs with three groups. We argue that the proper choice of method should be determined by the assessment which of the comparisons are considered primary and which are secondary, as determined by subject-matter considerations. We review four different methods that are simple to use with any standard software, but are substantially more powerful than frequently-used methods such as an ANOVA test followed by Tukey's method.*

**Introduction**

Medical research often involves comparing outcome measures between several experimental or observational groups. Designs with two groups are most common, and are well discussed in statistical handbooks. Multi-group designs are more complex, since they give rise to multiple possible between-group comparisons which are not always of equal interest to the researcher. If more than one such comparison is of interest, multiple testing problems arise. The proper statistical methods to deal with multiple comparisons are not exhaustively discussed in medical statistics handbooks. As a consequence, we see confusion among practitioners, and frequent use of suboptimal methods, such as analysis of variance (ANOVA) tests followed by Bonferroni, or by Tukey.

The reason that more powerful methods are not discussed in statistical handbooks is that such methods can become very complex in the general multi-group case, and such methods can often not be performed with standard commercial software. However, in the case of three groups improved methods are still relatively simple. In this paper we explain in detail the statistical issues involved in multiple comparisons with three groups. The three-group design is arguably the most frequently used of the multi-group designs, and this simplest of multi-group designs warrants a special discussion. It is especially in the three-group design that suboptimal methods are very frequently used in practice, while more powerful alternatives are available that are simple to use.

The subject of multiple has a long history and a huge literature, which we do not attempt to cover in full. This paper aims to give a practical overview of the subject for users, promoting methods that are both valid and powerful.

**Hypotheses for three groups**

Three-group experimental designs occur in many contexts, e.g. when comparing three categorical responses in a 2 by 3 table, or when comparing three numerical outcomes, parametrically or non-parametrically. We focus first on the standard example in which the outcome in the three groups is assumed to be normally distributed with equal variances in the groups, i.e. the classical ANOVA model. The means of the three groups are $\mu_1$, $\mu_2$, and $\mu_3$. There are four null hypotheses we may formulate about these means. First, the so-called 'global' null hypotheses that all means are equal:

$$H_{123}: \mu_1 = \mu_2 = \mu_3,$$

which we would normally test with an ANOVA *F*-test. Next, there are the three pairwise comparisons between groups:

$$H_{12}: \mu_1 = \mu_2; \qquad H_{13}: \mu_1 = \mu_3; \qquad H_{23}: \mu_2 = \mu_3,$$

which we would individually test with independent samples *t*-tests, or partial F-tests.

**Why correct for multiple comparisons?**

A false positive result in hypothesis testing is a rejection of a null hypothesis even though it is true. False positive results may result in an incorrect scientific finding reaching the scientific literature, and should therefore be prevented. The convention is to accept a probability of such a false positive result of at most 5%, which is achieved by requiring a p-value below 0.05. If multiple hypotheses are tested, each hypothesis again has a probability of 5% of a false positive result. Therefore, the probability that at least one a false positive result occurs as a result of the experiment will exceed the acceptable rate. In the three-group set-up (equal groups), simply testing all of the four hypotheses without correction results in an excessive 13% of experiments producing at least one false positive result if all hypotheses are true (see Supplemental Information). The corrections for multiple comparisons described below bring the probability of producing a false positive result back to the required 5% level. They can be used to guarantee that at least 95% of the three-group experiments performed produce no false positive results (Bretz, Hothorn, Westfall 2010, Hsu 1996, Hochberg and Tamhane 1987).

**Primary and secondary hypotheses**

The four hypotheses given above are seldom equally interesting to us, and we may usually distinguish between hypotheses of primary and secondary interest. Hypotheses of primary interest are those that are central to our research question. Hypotheses of secondary interest are those that are only of interest if we are able to reject one of the primary hypotheses. For example, if group 1 is the placebo group and groups 2 and 3 are two doses of a medicine, we may be primarily interested in hypotheses $H_{12}$ and $H_{13}$, comparing each dose of medicine to the placebo. The hypothesis $H_{23}$, that compares the two doses of medicine with each other, may become of interest only after we have shown that one (or both) of the doses works better than placebo. The hypothesis $H_{123}$ is not of interest at all in this case, as rejecting it only allows us to conclude that at least one of the doses is effective, but not which one.

Depending on the context of the experiment, different hypotheses may be primary and secondary. We distinguish four scenarios:

A. *The global hypothesis $H_{123}$ is primary*. This choice is implicit in the classical approach that starts with the overall ANOVA test, and digs deeper in the pairwise comparisons only if the ANOVA test was significant ('post hoc'). Generally, we should declare $H_{123}$ primary if rejection of $H_{123}$ alone leads to an interpretable result, even if none of the pairwise hypotheses would be rejected. An example is when the groups represent a continuous covariate, say BMI, cut into three clinically relevant categories, say underweight, normal weight and overweight. Rejection of $H_{123}$ in this case represents the presence of an association between covariate and outcome. The pairwise comparisons are of secondary interest.
B. *One pairwise hypothesis is primary,* say $H_{12}$. This approach is suitable if groups 1 and 2 are more important than group 3, or if the comparison between groups 1 and 2 is expected to have much more power than the other comparisons. For example, if group 1 is placebo, group 2 a treatment at high dose level, and group 3 the same treatment at reduced dose, we may be primarily interested in showing the effectiveness of the high dose. Only if the high dose is effective are we interested in showing the effectiveness of the lower dose.
C. *Two pairwise hypotheses are primary,* say $H_{12}$ and $H_{13}$. This approach is suitable if we are mainly interested in showing difference between group 1 and the other groups. For example, if group 1 is control and groups 2 and 3 are different experimental conditions, we may be mostly interested in showing that the two experimental conditions differ from control. Only if at least one of the experimental conditions differs from control are we interested in showing any difference between the two.

D. *All three pairwise comparisons are primary.* This approach is suitable if all three pairwise comparisons conditions are of equal interest, but rejection of the global null hypothesis $H_{123}$ is not meaningful by itself. For example, when comparing three different experimental conditions, all three pairwise comparisons may be equally relevant. Rejecting only $H_{123}$, however, would allow the conclusion that there some difference between the conditions somewhere, tantalizing but not informative.

The decision which hypotheses are primary and which are secondary should be based on subject-matter knowledge. It is important that this choice is made before seeing the data, or the protection offered by the multiple comparisons procedure is lost.

**Logical relationships between the hypotheses**

The four hypotheses $H_{123}$, $H_{12}$, $H_{13}$ and $H_{23}$ are closely related to each other. The following logical relationship between them holds: if any two are true, then all must be true. For example, if $H_{12}$ and $H_{13}$ are true, then $\mu_1 = \mu_2$ and $\mu_1 = \mu_3$, so $\mu_1 = \mu_2 = \mu_3$, which implies that $H_{23}$ and $H_{123}$ are also true. The number of true hypotheses can therefore be either 0, 1, or 4, but never 2 or 3. Additionally, if only one hypothesis is true, this cannot be $H_{123}$. These logical implications between the hypotheses play an important role in the powerful methods described below, as well as in subsequent interpretation of the results.

**Four procedures for four scenarios**

The choice of multiple testing method of choice depends on the division of the hypotheses into primary and secondary. We advocate four different methods for the four situations detailed above. All procedures follow two steps. In Step 1 all primary hypotheses are tested, using a proper multiple testing procedure if there is more than one primary hypothesis. If no primary hypothesis was rejected, the procedure stops. Otherwise, the secondary hypotheses are tested, and any primary hypotheses not rejected in Step 1 are tested again, simply at the 5% level. The procedures are as follows:

   A. $H_{123}$ *is primary (Closed).* In Step 1 we test $H_{123}$ at the 5% level. We proceed to Step 2 if $H_{123}$ is rejected. In Step 2 we test $H_{12}$, $H_{13}$ and $H_{23}$ each at the 5% level (Marcus, Peritz, Gabriel, 1976).
   B. $H_{12}$ *is primary (Shaffer).* In Step 1 we test $H_{12}$ at the 5% level. We proceed to Step 2 if $H_{12}$ is rejected. In Step 2 we reject $H_{123}$ immediately and test $H_{13}$ and $H_{23}$ each at the 5% level (Shaffer 1986).
   C. $H_{12}$ *and* $H_{13}$ *are primary (Step-down Dunnett).* In Step 1 we test $H_{12}$ and $H_{13}$ using Dunnett's procedure. We proceed to Step 2 if at least one of $H_{12}$ or $H_{13}$ is rejected. In Step 2 we reject $H_{123}$ immediately and test $H_{23}$ and any yet unrejected hypotheses from among $H_{12}$ and $H_{13}$, each at the 5% level (Dunnett 1955, Naik 1975).
   D. $H_{12}$, $H_{13}$ *and* $H_{23}$ *are primary (Step-down Tukey).* In Step 1 we test $H_{12}$, $H_{13}$ and $H_{23}$ using Tukey's procedure. We proceed to Step 2 if at least one hypothesis is rejected. In Step 2 we we reject $H_{123}$ immediately, and test any yet unrejected hypotheses from among $H_{12}$, $H_{13}$ and $H_{23}$ again, each at the 5% level (Tukey 1953, Finner 1988).

These procedures are easy to perform. Dunnett's and Tukey's procedures, and their variants for unequal-size groups (Spurrier and Isham 1985, Kramer 1956, Dunnett and Tamhane 1991) are available in all standard statistical software packages.

**Why are these procedures valid?**

The validity of these stepwise procedures may be understood using either closed testing (Marcus, Peritz and Gabriel 1976) or the Sequential Rejection Principle (Goeman and Solari 2010). Loosely speaking, the latter principle says that we may control for multiple testing using stepwise procedures, and that every step after the first may assume that all rejections made in previous steps were correct (since otherwise a false positive result was already obtained). In the three-group problem, this means that in the second step we may assume that at least one hypothesis is false. By the logical implications, this means that at most one hypothesis may be true. If there is only one true hypothesis, all tests may be done at the 5% level, since there is no multiple testing issue left. For this reason, after rejecting at least one primary hypothesis, all remaining hypotheses may be tested simply at the 5% level. Additionally, after rejecting any of the primary hypotheses we may immediately reject $H_{123}$, since that hypothesis must be false if at least one hypothesis is false.

**Power**

Although the choice of method should primarily depend on the relative order of priority of the hypotheses, power can also be an important consideration. Comparing the different methods, a rule of thumb is that the method is most powerful that declares as primary the hypotheses for which the test has most power. This is because the power for any hypotheses is bounded by the power for rejecting the primary hypothesis. It follows, for example, that Shaffer's method for case B is relatively powerful if we have reason to believe that the means of Groups 1 and 2 are furthest apart, or if there are much fewer subjects in group 3. Step-down Dunnett is relatively powerful if the group mean of Group 1 is most outlying, or if Group 1 has more subjects than the other groups.

Although it may not appear so at first sight, the Closed method and Tukey's step-down method have very similar power in practice. This also holds if we look only at power for rejecting the pairwise hypotheses $H_{12}$, $H_{13}$ and $H_{23}$ relevant in scenario D. For example, with a balanced design with 6 subjects per group, $\sigma^2 = 1$, and true means $\mu_1 = 1$, $\mu_2 = 0$ and $\mu_3 = -1$, Tukey's step-down method and the Closed method give exactly the same rejections 97.4% of the time. Tukey's method is slightly more powerful when only the pairwise hypotheses are considered, but the power difference is minimal: 80.8% for Tukey versus 80.6% for Closed. Closed testing is therefore an attractive alternative in Scenario D because of its simplicity. This will become relevant when we turn to other models than the simple ANOVA-model case.

Methods frequently advocated and used in the literature include an ANOVA test followed by a single-step Tukey correction for the pairwise hypotheses, or ANOVA followed by Bonferroni correction. Both Closed testing and Step-down Tukey are uniformly more powerful than such approaches, at least in a balanced design, implying that they never give fewer rejections. Often the number of rejections is much greater. For example, in the balanced design example considered above, the probability that Tukey's step-down method rejects a larger number of pairwise hypotheses than ANOVA followed by single-step Tukey is 26%, and the same method rejects more than ANOVA followed by Bonferroni with probability 31%. Comparable probabilities can be calculated for the Closed procedure.

**Adjusted p-values**

As described above, the procedures only tell whether to reject or not to reject at the fixed multiple significance level of 5%. Often, however, it is desirable to report multiplicity-adjusted p-values for each hypothesis, to assess the strength of the evidence against that hypothesis. Multiplicity-adjusted p-values have an easy interpretation: if it is smaller than 5%, this means that the corresponding hypothesis is rejected by the multiple testing procedure at the 5% level. The following table allows quick calculation of the adjusted p-values for the four procedures discussed above. Here $p_{123}$, $p_{12}$, $p_{13}$ and $p_{23}$ are the p-values of the tests of $H_{123}$, $H_{12}$, $H_{13}$ and $H_{23}$, respectively. We denote the Tukey-adjusted p-values, which can be obtained from standard software, by $\tilde{p}_{12}^T$, $\tilde{p}_{13}^T$ and $\tilde{p}_{23}^T$, and define

$$\tilde{p}^T = \min(\tilde{p}_{12}^T, \tilde{p}_{13}^T, \tilde{p}_{23}^T)$$

as the smallest among them. Similarly, we denote the Dunnett-adjusted p-values by $\tilde{p}_{12}^D$ and $\tilde{p}_{13}^D$, and

$$\tilde{p}^D = \min(\tilde{p}_{12}^D, \tilde{p}_{13}^D)$$

as the smallest among those.

Table 1 describes the calculation of adjusted p-values for the four scenarios. It simply involves taking the maximum between the p-value of the hypothesis itself and the best adjusted p-value from the primary hypotheses in Step 1.

Table 1: Adjusted p-values for the four hypotheses in the four scenarios.

| Scenario | Method | $H_{12}$ | $H_{13}$ | $H_{23}$ | $H_{123}$ |
|---|---|---|---|---|---|
| A | Closed | $\max(p_{12}, p_{123})$ | $\max(p_{13}, p_{123})$ | $\max(p_{23}, p_{123})$ | $p_{123}$ |
| B | Shaffer | $p_{12}$ | $\max(p_{13}, p_{12})$ | $\max(p_{23}, p_{12})$ | $p_{12}$ |
| C | Step-down Dunnett | $\max(p_{12}, \tilde{p}^D)$ | $\max(p_{13}, \tilde{p}^D)$ | $\max(p_{23}, \tilde{p}^D)$ | $\tilde{p}^D$ |
| D | Step-down Tukey | $\max(p_{12}, \tilde{p}^T)$ | $\max(p_{13}, \tilde{p}^T)$ | $\max(p_{23}, \tilde{p}^T)$ | $\tilde{p}^T$ |

For completeness, in Table 1 we have given adjusted p-values for $H_{123}$ also in scenarios B, C and D, although usually, in these scenarios, researchers will generally not be very much interested in that hypothesis. In these scenarios this adjusted p-value can still be interpreted as the adjusted significance for rejecting $H_{123}$ even though its calculation does not involve the ANOVA p-value $p_{123}$.

**Example**

Let us illustrate the different methods with a more concrete example. Suppose a small-scale randomized clinical trial has been performed comparing placebo and two doses of a medicine (low dose and high dose) in equal groups of 20 subjects. The mean response in the three groups was 11.5 (group 1: placebo), 12.8 (group 2: low dose) and 14.1 (group 3: high dose), with an estimated common standard deviation of 1.9. The data are visualized in Figure 1.

Figure 1: Visualisation of the example data: response distribution in the three groups.

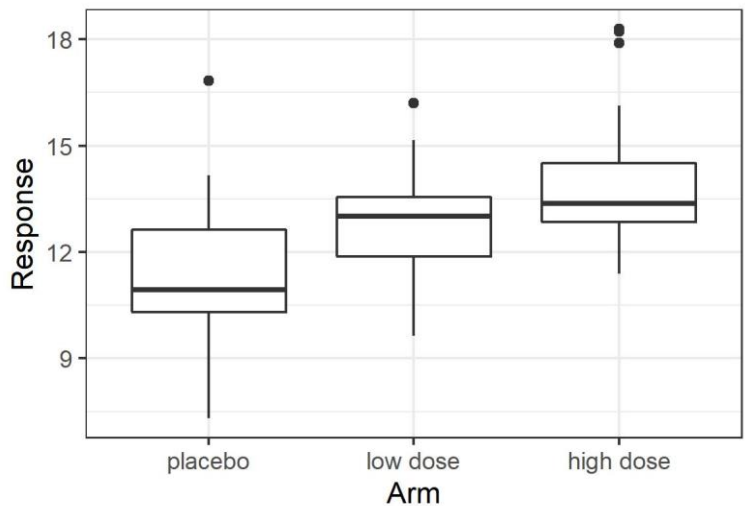

To assess significance of differences between the mean response in the groups, we can calculate the p-values for the three comparisons as $p_{12} = 0.027$, $p_{13} < 0.001$, $p_{23} = 0.037$, but we should not interpret these p-values directly without applying a multiple comparisons procedure first.

The suboptimal approaches of an ANOVA test followed by Tukey or Bonferroni will reject $H_{123}$ since the ANOVA p-value is $p_{123} < 0.001$, and also $H_{13}$, but not $H_{12}$ or $H_{23}$, since the Tukey-adjusted p-values for these hypotheses are 0.068 and 0.092, respectively, and the Bonferroni-adjusted p-values are even higher. We note that this result is logically incomplete: if $H_{13}$ is false, then we know that at least one of $H_{12}$ or $H_{23}$ must be false, but we cannot confidently say which one.

The choice of the proper multiple testing method depends on the context. Closed testing (A) is appropriate if we would first and foremost want to show that there is some effect of the treatment, regardless of dose. Shaffer's method (B) prioritizing $H_{12}$ would be most appropriate if demonstrating the effectiveness of the low dose would be of primary interest, e.g. if high dose was expected to have many side effects. Dunnett's method would be appropriate if we would be primarily interested in finding at least one of the doses different from placebo. Tukey's method would be chosen if we would be equally interested in showing a difference between any of the groups, but if only rejecting the global hypothesis would be unsatisfactory.

The adjusted p-values for the four hypotheses for the four methods are given in Table 2, after calculating that the smallest adjusted Dunnett- and Tukey adjusted p-values are both < 0.001. With all four methods we may claim significant differences between all three treatment groups. Shaffer's method B shows less confidence than the other methods because it prioritized a hypothesis with a relatively small difference.

*Table 2: Adjusted p-values for the four hypotheses in the four scenarios in the example.*

| Scenario | *Method* | $H_{12}$ | $H_{13}$ | $H_{23}$ | $H_{123}$ |
|---|---|---|---|---|---|
| A | Closed | 0.027 | < 0.001 | 0.037 | < 0.001 |
| B | Shaffer | 0.027 | 0.027 | 0.037 | 0.027 |
| C | Step-down Dunnett | 0.027 | < 0.001 | 0.037 | < 0.001 |
| D | Step-down Tukey | 0.027 | < 0.001 | 0.037 | < 0.001 |

**Paradoxical outcomes**

The logical relationships between the hypotheses dictate that the number of hypotheses may be 0, 1 or 4, but never 2 or 3. The result of the test procedure, however, may not always conform to this. In particular it may happen that only one of $H_{12}$, $H_{13}$ or $H_{23}$ is rejected. In such cases we may significantly claim that at least one more hypothesis is false, only we are not sure which one. In scenario A the frustrating event may even occur that $H_{123}$ is rejected, but none of $H_{12}$, $H_{13}$ and $H_{23}$. Then we may significantly claim that at least two more hypotheses are false, only that we are not confident which ones. However, this event is much rarer with the procedures escribed in this paper than with some of the suboptimal procedures that have been frequently used in practice, such as ANOVA followed by Bonferroni (Sedgwick 2014).

**Other models**

Three-group comparisons occur in many more contexts than the ANOVA model we have discussed so far, for example when comparing three proportions using chi-squared tests in a 2 × 3 table, when performing non-parametric analysis with Kruskal-Wallis test, when comparing three survival curves using a log-rank test, or in regression models when considering a categorical covariate with three levels. In all such cases we can formulate a global null hypothesis $H_{123}$ of equality of all three groups and corresponding pairwise hypotheses $H_{12}$, $H_{13}$ and $H_{23}$. Regardless of the model considered, the logical relationships between hypotheses and the potential division of the hypotheses into primary and secondary remain the same, so that the same four scenarios arise.

Practically, however, there is a difference between the ANOVA model context and other models. While the Closed and Shaffer methods (A and B) may still be used by simply applying them to *p*-values from model-appropriate tests, analogues of Dunnett's and Tukey's methods are generally unavailable in commercial statistical software packages, even though the statistical theory to perform them is available and implemented in R (Hothorn, Bretz, Westfall 2008). In case such methods are out of reach of practitioners we recommend the method for case A as a practical alternative to use in scenarios C and D.

**Four or more groups and other extensions**

While the three-group case is relatively simple, complexity quickly explodes as the number of groups increases. With four groups, there are already 6 pairwise null hypotheses and 4 three-group null hypotheses, aside from the global null hypothesis. With so many hypotheses there are exceedingly many possible divisions into primary, secondary and perhaps tertiary hypothesis, and therefore as many possible methods. Moreover, the complexity of methods also increases, and powerful methods may easily involve three of more steps. Still, the same principles apply as in the three-group case. Extensions of all four methods exist (Bretz, Hothorn, Westfall 2010), and the general closed testing (Marcus, Peritz and Gabriel 1976) or sequential rejection principles (Goeman and Solari 2010) may be used as a valuable guiding principle. We leave this subject to the specialized literature, and recommend that interested users consult a statistician.

Simultaneous confidence intervals for the pairwise differences may also be calculated. Such calculations are relatively easy for single-step procedures, but more involved for the step-down variants (Dmitrienko, Tamhane, Bretz 2009).

**Conclusion**

When comparing three groups, multiple testing issues come into play, but methods are still relatively simple. We have reviewed four simple and powerful methods that, surprisingly enough are not frequently used. We have clarified which of the methods to choose in which situation, depending on a content-driven division of the hypotheses into primary and secondary. Use of these methods will increase power compared to the use of frequently used suboptimal methods, such as an ANOVA test followed by single step Tukey. The use of more powerful methods for multiple testing correction may increase the acceptance and use of multiple testing methodology, which is crucial for prevention of false positive results in science.

**Reproducibility**

The code in R for all sumulations and analysis is given in the supplemental information.